\documentclass[letterpaper]{article} 
\usepackage{aaai24}  
\usepackage{times}  
\usepackage{helvet}  
\usepackage{courier}  
\usepackage[hyphens]{url}  
\usepackage{graphicx} 
\usepackage{natbib}  
\usepackage{caption} 
\usepackage{algorithm}
\usepackage{algorithmic}

\usepackage{newfloat}
\usepackage{listings}
\DeclareCaptionStyle{ruled}{labelfont=normalfont,labelsep=colon,strut=off} 
\floatstyle{ruled}
\newfloat{listing}{tb}{lst}{}
\floatname{listing}{Listing}
\title{\textbf{\textsc{\textsf{VITA}}}: ‘Carefully Chosen and Weighted Less’ Is Better \\ in Medication Recommendation}
\author{
    Taeri Kim\textsuperscript{\rm 1}\equalcontrib,
    Jiho Heo\textsuperscript{\rm 1}\equalcontrib,
    Hongil Kim\textsuperscript{\rm 2}\equalcontrib,
    Kijung Shin\textsuperscript{\rm 3}\thanks{Co-corresponding authors.}, 
    Sang-Wook Kim\textsuperscript{\rm 1}\footnotemark[2]
}
\affiliations{
    \textsuperscript{\rm 1}Department of Computer Science, Hanyang University, South Korea\\
    \textsuperscript{\rm 2}Department of Artificial Intelligence, Hanyang University, South Korea\\
    \textsuperscript{\rm 3}Kim Jaechul Graduate School of AI \& School of Electrical Engineering, KAIST, South Korea\\
    \{taerik, linda0123, hong0814, wook\}@hanyang.ac.kr, kijungs@kaist.ac.kr
}

\usepackage{multirow}
\usepackage{booktabs}
\usepackage{amsmath}
\usepackage{amssymb}

\begin{document}

\maketitle

\begin{abstract}
We address the medication recommendation problem, which aims to recommend effective medications for a patient's current visit by utilizing information ({\it e.g.}, diagnoses and procedures) given at the patient's current and past visits.
While there exist a number of recommender systems designed for this problem, we point out that they are challenged in accurately capturing the relation ({\it spec.}, the degree of relevance) between the current and each of the past visits for the patient when obtaining her {\it current health status}, which is the basis for recommending medications.
To address this limitation, we propose a novel medication recommendation framework, named \textbf{\textsc{\textsf{VITA}}}, based on the following two novel ideas: (1) relevant-\underline{\textbf{V}}isit select\underline{\textbf{I}}on; (2) \underline{\textbf{T}}arget-aware \underline{\textbf{A}}ttention. Through extensive experiments using real-world datasets, we demonstrate the superiority of {\textsc{\textsf{VITA}}} ({\it spec.}, up to 5.56\% higher accuracy, in terms of Jaccard, than the best competitor) and the effectiveness of its two core ideas.
The code is available at \url{https://github.com/jhheo0123/VITA}.
\end{abstract}

\section{Introduction}\label{sec:introduction}

{\it Medication recommendation} aims to help doctors in prescribing effective medications for a patient's current visit~\cite{leap, gamenet}.
When prescribing medications to a patient, doctors should consider the following factors:
(1) the diagnoses and procedures given to her at the current visit;
(2) her past health records;
(3) the relations between medications to be prescribed ({\it e.g.}, the possibility of adverse effects when taken together).
This is a time-consuming and challenging process even for experienced doctors~\cite{cognet}.
Therefore, the medication recommendation that can alleviate these difficulties of doctors has been an important research area in medical tasks.

Early medication recommender systems recommend medications to a patient by considering {\it only} the diagnoses and procedures given to her at the current visit ({\it i.e.}, {\it current visit information}), which are called {\it instance-based} methods~\cite{leap, ppc, smr}. However, they overlooked the fact that, although patients are given with the same diagnoses at the current visit, the cause of the given diagnoses may be different for each patient, so the medications for them could also be different~\cite{gamenet, safedrug}; this is because they do not utilize the patients' past health records~\cite{cognet}.

To alleviate this limitation, {\it longitudinal-based} medication recommender systems have emerged, which utilize not only a patient's {\it current} visit information but also her {\it past} health records~\cite{gbert,gamenet,premier,sarmr,mrsc, safedrug,csedrug,cognet}.
These methods usually consist of an {\it encoder} and a {\it predictor}.
For their encoders, most of them~\cite{gamenet,premier,sarmr,mrsc,safedrug,csedrug} learn a {\it patient representation} indicative of {\it her current health status} by aggregating her current visit information and the diagnoses and procedures given at her past visits  ({\it i.e.}, {\it past visit information}) in consideration of the visit order via a Recurrent Neural Network (RNN)-based model.
Then, for their predictors, they recommend medications to her by utilizing other past health records ({\it e.g.}, medications prescribed at her past visits) based on their relevance with her patient representation via various deep-learning models ({\it e.g.}, an attention network~\cite{attention}).
On the other hand, COGNet~\cite{cognet}, another longitudinal-based method recently proposed, encodes {\it only} a patient’s {\it current} visit information, unlike the above methods, to obtain her current health status in the encoder by capturing the relation between diagnoses (resp. procedures) at her current visit via a transformer~\cite{transformer}-based model.
Then, in the predictor, COGNet additionally takes into account the similarity between (a) medications that should be prescribed at her current visit by considering only her current visit information and (b) those prescribed at her past visits, to improve upon the predictor of the aforementioned methods. Please refer to the Appendix\footnote{The Appendix for {\textsc{\textsf{VITA}}} can be found at \url{https://github.com/jhheo0123/VITA}.} for more detailed information on these ({\it i.e.}, related studies).

Although longitudinal-based methods achieve higher accuracy than instance-based methods by utilizing a patient's past health records, we point out a key limitation of them:
when {\it obtaining the patient's current health status} ({\it i.e.}, in the {\it encoder}), they have a difficulty in accurately capturing the relation ({\it spec.}, the degree of relevance) of each past visit to the current one.
Specifically, they either compute the relevance score by considering simply the {\it order of visits} via a RNN-based model~\cite{gamenet,premier,sarmr,mrsc,safedrug,csedrug}, or do not attempt to capture the degree of relevance, not taking into account {\it any} of the patient's past visits~\cite{cognet}.
However, when obtaining the patient's current health status, accurately capturing the degree of relevance between her current and past visits is highly important.
In particular, her past visit information relevant (resp. irrelevant) to her current visit should (resp. should not) be reflected in representing her current health status.
Our empirical findings supporting this claim are detailed in Section~\ref{sec: motivation}.

In this paper, we propose {\textsc{\textsf{{VITA}}}} (relevant-visit selection and target-aware attention), a novel medication recommendation framework, based on a more accurate understanding of her current health status.
To achieve this, we learn a patient representation indicative of her current health status by taking into account an accurate relevance score between the current visit and each of the past ones, without relying solely on the visit order.
Furthermore, we do {\it not} consider her past visits {\it irrelevant} to her current one {\it at all}, instead of simply assigning lower relevance scores to them, since this approach enhances accurate medication recommendation.
Our empirical validation of these ideas can be found in Section~\ref{sec:evaluation}.

Our contributions are summarized as follows:
\begin{itemize}
\item \textbf{Important Discovery}: We discovered that existing medication recommender systems face challenges in accurately capturing the degree of relevance of each past visit to the current one when obtaining a patient’s current health status. In addition, we are the first to demonstrate that using past visits irrelevant to the current one has a negative effect on recommending accurate medications.
\item \textbf{Novel Framework}: To address this key limitation, we propose a novel medication recommendation framework, named \textbf{{\textsc{\textsf{{VITA}}}}}, that recommends medications by employing an {\it enhanced patient representation}, based on the following two novel ideas: (1) relevant-\underline{\textbf{V}}isit select\underline{\textbf{I}}on that automatically excludes past visits irrelevant to the current one; (2) \underline{\textbf{T}}arget-aware \underline{\textbf{A}}ttention that accurately captures the relevance score of the past visits to the current one.
\item \textbf{Extensive Evaluation}: We validate the effectiveness of {\textsc{\textsf{{VITA}}}} through extensive experiments using public real-world datasets. Most importantly, {\textsc{\textsf{{VITA}}}} surpasses {\it all} six state-of-the-art competitors, achieving an improvement of up to 5.56\% in terms of Jaccard.
Also, the two core ideas of {\textsc{\textsf{{VITA}}}} can be orthogonally combined with existing medication recommender systems; when combined, they elevate the accuracy beyond that of the original ones.
\end{itemize}

\begin{figure}[t]
\centering
\includegraphics[width=0.5\textwidth]{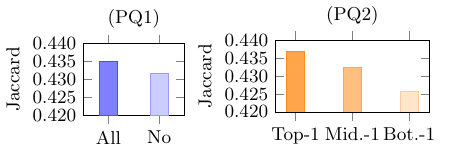}
\caption{Accuracies of GAMENet when varying the use of past visit information of a patient. `All' (resp. `No') means to use all (resp. not to use any) past visit information. `Top-1/Mid.-1/Bot.-1' means to use only one past visit information that is the most/moderately/the least similar to the current visit information, respectively. All differences are statistically significant with a $p$-value $\leq 0.001$.}\label{fig:motivation}
\end{figure}

\section{Motivation}\label{sec: motivation}
In this section, we demonstrate the limitations of existing medication recommender systems via preliminary experiments answering the following preliminary questions (PQs):

\begin{itemize} 
\item \textbf{PQ1}: Does using past visit information in representing a patient's current health status help accurate medication recommendation?
\item \textbf{PQ2}: Which past visit information is beneficial to better representing a patient’s current health status?
\end{itemize}

\noindent\textbf{Experimental Settings}. We conducted experiments using the GAMENet model~\cite{gamenet}\footnote{The results on other existing medication recommender systems showed similar tendencies to those on the GAMENet. These results would be shown in the Appendix.}, the MIMIC-III dataset~\cite{mimic}, and the accuracy measure Jaccard, which are most commonly used in medication recommendation~\cite{premier,sarmr, mrsc, micron, safedrug, cognet}.
Please refer to Section~\ref{sec:evaluation} for more detailed information on experimental settings.

\begin{figure*}[t]
\centering
\includegraphics[width=0.99\textwidth]{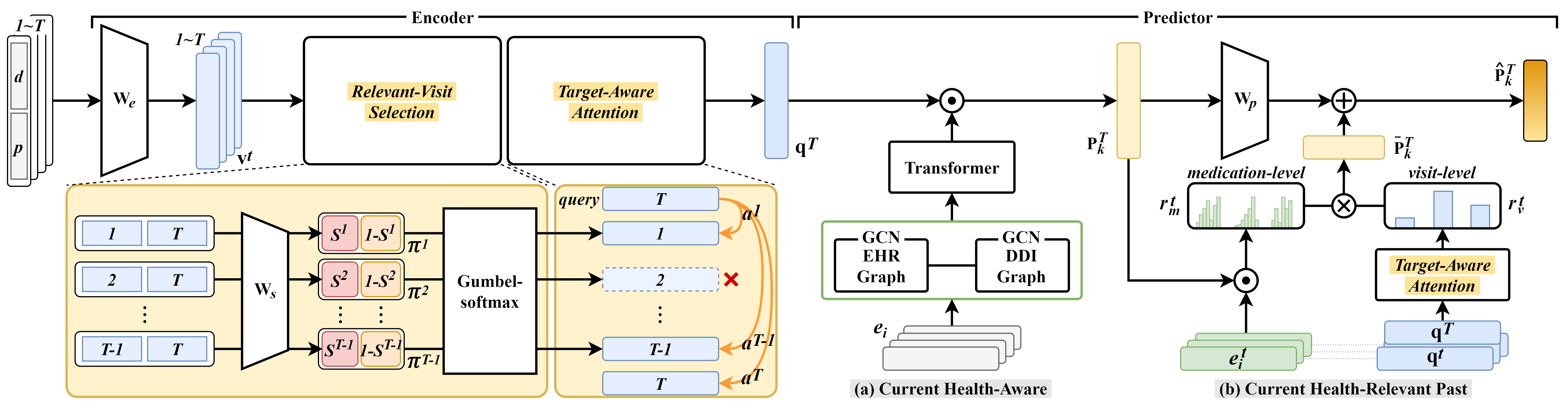}
\caption{Overview of {\textsc{\textsf{{VITA}}}}, composed of two components: an encoder based on relevant-visit selection and target-aware attention; a predictor based on (a) current health-aware and (b) current health-relevant past medication representations.} \label{fig:vita}
\end{figure*}

\noindent\textbf{PQ1: Effect of the use of a patient's past visit information}. We first analyze whether it is really helpful to use a patient's past visit information for representing her current health status ({\it i.e.}, in her patient representation) for recommending medications at the current visit.
To this end, we compared the accuracy of GAMENet that uses {\it all} past and current visit information as input to the RNN-based model in the encoder ({\it i.e.}, the original method) with that of its counterpart that uses {\it only} her {\it current} visit information.

The results for the above two methods are represented as `All' and `No', respectively, in Figure~\ref{fig:motivation}-(PQ1). From the results, we found that using not only the current visit information but also the past ones for patient representation is helpful for accurate medication recommendation.

\noindent\textbf{PQ2: Effect of the use of a patient's past visit information depending on its degree of relevance to her current visit}. Now, we analyze which past visit information of a patient is helpful to obtain better patient representation in medication recommendation. To this end, we first calculated the Jaccard similarity between the current and each of past visit information for each patient.
Specifically, we first represented the diagnoses (resp. procedures) given at each visit as a multi-hot vector for each visit per patient.
Then, we represented each visit as a single vector by concatenating the two multi-hot vectors of the diagnoses and procedures to calculate the similarity. After that, for each patient, we measured the accuracy of GAMENet using {\it only} the past visit information {\it most similar} to the current visit, along with the current one, as an input to the RNN-based model in the encoder. The result of this method is represented as `Top-1' in Figure~\ref{fig:motivation}-(PQ2).
Also, for comparison, we additionally measured the accuracy using moderately (resp. the least) similar one, {\it instead of} using the most similar past visit information (the moderately similar past visit is the one with the median similarity).
The result for this method is represented as `Mid.-1' (resp. `Bot.-1') in Figure~\ref{fig:motivation}-(PQ2).

As shown in Figure~\ref{fig:motivation}-(PQ2), the accuracy decreases in the order of using only the most similar, moderately similar, and the least similar past visit information along with the current one.
In addition, we found that the accuracy when using {\it only} the most similar (resp. the least similar) past visit information along with the current one is rather {\it higher} (resp. {\it lower}) than that of the original method using {\it all} past visit information along with the current one (resp. the method using {\it only} the current visit information) (compare `All' (resp. `No') and `Top-1' (resp. `Bot.-1') accuracies in Figure~\ref{fig:motivation}-(PQ1) and -(PQ2), respectively).

\noindent\textbf{Summary}. Based on the above results, we draw the following conclusions: (1) referencing past visit information in obtaining the patient representation is {\it helpful} for medication recommendation; furthermore, (2) using {\it only} relevant ({\it i.e.}, similar) past visit information to the current one is {\it more} helpful; however, (3) using irrelevant ({\it i.e.}, dissimilar) past visit information to the current one has a {\it negative} effect.

\section{\textsc{\textsf{{VITA}}}: Proposed Framework}\label{sec:approach}

In this section, we detail our proposed framework, {\textsc{\textsf{{VITA}}}}, based on relevant-visit selection and target-aware attention.

\subsection{Problem Definition}\label{sec:definition}
\noindent\textbf{Definition 1: Patient Records}. In Electronic Health Records (EHR) data ({\it e.g.}, MIMIC-III dataset), the health records of each patient $x$ consist of sequential visits $\mathcal{V}_x=\ [\mathbf{V}_x^1,\ \cdots, \mathbf{V}_x^{\left(T-1\right)},\ \mathbf{V}_x^T]$, where $\mathbf{V}_x^{(T-1)}$ denotes the $(T-1)$-th visit of patient $x$. For simplicity, following~\cite{leap,gbert,gamenet,cognet}, we omit the subscript indicating a patient ({\it i.e.}, $x$) and describe {\textsc{\textsf{{VITA}}}} with a single patient. Therefore, the patient’s visits are denoted by $\mathcal{V}=[\mathbf{V}^1,\ \cdots,\ \mathbf{V}^{(T-1)},\ \mathbf{V}^T]$. Each visit $\mathbf{V}^{(T-1)}$ of a patient consists of three subsets of all diagnoses $\mathcal{D}$, all procedures $\mathcal{P}$, and all medications $\mathcal{M}$; the subsets are represented as multi-hot vectors ({\it e.g.}, $\mathbf{d}^{(T-1)}\in\mathbb{R}^{|\mathcal{D}|}, \mathbf{p}^{(T-1)}\in\mathbb{R}^{|\mathcal{P}|}$, and $\mathbf{m}^{(T-1)}\in\mathbb{R}^{|\mathcal{M}|}$).

\noindent\textbf{Definition 2: EHR and DDI Graphs}.
The EHR and the Drug-Drug Interactions (DDI) graphs are denoted by $\mathcal{G}_{EHR}=(\mathcal{M}, \mathcal{E}_{EHR})$ and $\mathcal{G}_{DDI}=(\mathcal{M}, \mathcal{E}_{DDI})$, respectively, where $\mathcal{E}_{EHR}$ denotes the set of edges between medications, each of which indicates that two medications have been prescribed together at any visit of any patient, and $\mathcal{E}_{DDI}$ denotes the set of edges between medications, each of which indicates that two medications may cause adverse effects if taken together; their adjacency matrices $\mathbf{A}_{EHR}$, $\mathbf{A}_{DDI}\in\mathbb{R}^{|\mathcal{M}|\times|\mathcal{M}|}$ satisfy: $\mathbf{A}_{EHR}[i,j] = 1$ if and only if the medications $i$ and $j$ have been prescribed together at any visit of any patient, $\mathbf{A}_{DDI}\ [i,\ j] = 1$  if and only if the medication $i$ and $j$ can be harmful when they are taken together. The same EHR and DDI graphs are used for all patients.

\noindent\textbf{Medication Recommendation Problem}. Given past health records $[\mathbf{V}^1,\ \cdots,\ \mathbf{V}^{(T-1)}]$ and current visit information ({\it i.e.}, diagnoses $\mathbf{d}^T$ and procedures $\mathbf{p}^T$ given at the current visit $T$) of a patient, and EHR and DDI graphs, the goal is to recommend the medications $\hat{\mathcal{M}}^{T}$ for her current visit $T$. Key notations used in this paper can be found in the Appendix.

\subsection{Key Components in {\textsc{\textsf{{VITA}}}}}\label{sec:key_component}
{\textsc{\textsf{{VITA}}}} consists of two components, an encoder and a predictor, and its schematic overview is presented in Figure~\ref{fig:vita}. In the following, we delve into the details of each component. 

\noindent\textbf{Encoder}.
In its encoder, {\textsc{\textsf{{VITA}}}} aims to obtain an enhanced patient representation $\mathbf{q}^T\in\mathbb{R}^{dim}$, which denotes her current health status ({\it i.e.}, at the $T$-th visit), by employing relevant-visit selection and target-aware attention.
To achieve this, for each visit, {\textsc{\textsf{{VITA}}}} first concatenates visit $t$'s diagnoses $\mathbf{d}^t$ and procedures $\mathbf{p}^t$, and then feeds it into an embedding layer to obtain a dense representation $\mathbf{v}^t$ of visit $t$, as follows:
$\forall t \in \{1,\cdots, (T-1), T\}$,
\small
\begin{equation}\label{eq:visit_embedding}
    \begin{split}
    \mathbf{v}^t = {concat(\mathbf{d}^t,\mathbf{p}^t)}\mathbf{W}_e,
    \end{split}
\end{equation}
\normalsize
where $\mathbf{W}_e\in\mathbb{R}^{(|\mathcal{D}|+|\mathcal{P}|)\times dim}$ denotes the learnable weight matrix of the embedding layer.

\noindent\textbf{\textit{Relevant-Visit Selection}}.
Then, {\textsc{\textsf{{VITA}}}} aggregates the dense representations $\mathbf{v}^t$ for all visits to obtain her patient representation $\mathbf{q}^T$.
Recall that, in Section~\ref{sec: motivation}, we showed that using {\it only} the past visit information {\it relevant} to the current one is helpful for accurate medication recommendation. Therefore, we design a novel {\it relevant-visit selection} module to find only the past visits relevant to the current visit. However, here we encounter the following challenge: the number of past visits relevant to the current visit may {\it differ per patient}; for some patients, it is possible that {\it no} past visit is relevant to the current visit. To address this challenge, we predict per patient whether each of the past visits is relevant to her current one or not ({\it i.e.}, whether to select or not).

Specifically, {\textsc{\textsf{{VITA}}}} obtains a set $\mathcal{V}^{rel}$ of past visits relevant to the current visit via the relevant-visit selection module.
To this end, for each visit, {\textsc{\textsf{{VITA}}}} first concatenates the dense representations $\mathbf{v}^t$ of the past visit $t$ and $\mathbf{v}^T$ of the current visit $T$ to provide the relevant-visit selection module with the context for deciding whether to select the past visit $t$, and then feeds them into a Multi-Layer Perceptron (MLP) layer to obtain the probability $s^t$ of the past visit $t$ being selected, as follows:
$\forall t \in \{1, \cdots, (T-1)\}$,
\small
\begin{equation}\label{eq:select_embedding}
    \begin{split}
    {s}^t = sigmoid({concat(\mathbf{v}^t,\mathbf{v}^T)}\mathbf{W}_s+b_s),
    \end{split}    
\end{equation}
\normalsize
where $\mathbf{W}_s\in\mathbb{R}^{2dim\times1}$ and $b_s \in \mathbb{R}$ denote a weight matrix and a bias value of the MLP layer, respectively.
Finally, {\textsc{\textsf{{VITA}}}} employs the Gumbel-softmax~\cite{gumbel1}, which transforms the discrete sampling problem into a differentiable continuous problem, thereby allowing for the flow of gradients, to select the past visits relevant to the current visit ({\it i.e.}, to obtain the set $\mathcal{V}^{rel}$), as follows:
\small
\begin{align}\label{eq:gumbel_softmax}
    \mathcal{V}^{rel} = \{\mathbf{v}^t : t \in \{1, \cdots, (T-1)\} \text{ and } \lfloor o_1^t + 0.5 \rfloor = 1\},
\end{align}
\normalsize
where $o_{\gamma}^t =\frac{exp((\text{log}\pi_\gamma^t+z_\gamma^t)/\tau_{g})}{\sum_{{\delta}=1}^{2}exp((\text{log}\pi_\delta^t+z_\delta^t)/\tau_{g}))}$, and  $\pi^t\in\mathbb{R}^{2\times1}$ denotes a vector obtained by concatenating the probability $s^t$ and the probability $(1-s^t)$ of each past visit $t$ not being selected; $z_\gamma^t$ and $\tau_{g}$ denote the $\gamma$-th sample from the $Gumbel(0,1)$ distribution for the past visit $t$ and the temperature hyperparameter, respectively.

\noindent\textbf{\textit{Target-Aware Attention}}.
Given the set $\mathcal{V}^{rel}$, {\textsc{\textsf{{VITA}}}} aggregates the dense representations $\mathbf{v}^t$ in $\mathcal{V}^{rel}$ and  $\mathbf{v}^T$ of the current visit $T$ to  obtain the ultimate patient representation $\mathbf{q}^T$.
Here, to the best of our knowledge, we are first to employ an attention network in the {\it encoder} of medication recommender systems; this aims to accurately capture the relevance score $\alpha^t$ between the current visit $T$ and each relevant past visit  $t$, without relying solely on the visit order\footnote{Nevertheless, we experimented with a variant of {\textsc{\textsf{{VITA}}}} that incorporates positional encoding into our target-aware attention, but the current version of {\textsc{\textsf{{VITA}}}} showed higher accuracy.}.
Moreover, to obtain an accurate relevance score $\alpha^t$ even when {\it all} the past visits are {\it weakly} relevant to the current visit, we design a novel {\it target-aware attention} module that assigns the relevance score ${\alpha}^t$ of each past visit $t$ relatively not only to the other past visit but also to the current visit.

Specifically, using the dense representation $\mathbf{v}^T$ of the current visit $T$ as a query and all dense representations $\mathbf{v}^t\in\mathcal{V}^{rel}$ for the relevant past visits $t$ and $\mathbf{v}^T$ of the current visit $T$ as keys and values, {\textsc{\textsf{{VITA}}}} computes the relevance score $\alpha^t$ between the current visit and each of relevant past visits, as follows: $\forall \mathbf{v}^t\in \mathcal{V}^{rel}\cup{\{\mathbf{v}^T\}},$
\small
\begin{equation}\label{eq:relevance_score}
    \begin{split}
    \alpha^t = \frac{exp((\mathbf{v}^T\mathbf{W}_\alpha{\mathbf{v}^t}^{\intercal})/\sqrt{dim})}{\sum_{\mathbf{v}^f\in \mathcal{V}^{rel}\cup{\{\mathbf{v}^T\}}}exp((\mathbf{v}^T\mathbf{W}_\alpha{\mathbf{v}^f}^{\intercal})/\sqrt{dim})},
    \end{split}
\end{equation}
\normalsize
where $\mathbf{W}_\alpha\in\mathbb{R}^{dim\times dim}$ denotes a learnable weight matrix of the target-aware attention module.
It is worth noting that the relevance score $\alpha^T$ for the current visit $T$ is also calculated; thus, the sum of the softmax values for all selected past visits (except for the current visit) is not necessarily fixed to 1; it is also possible for the sum to be much smaller than 1 if {\it all} selected visits are {\it weakly relevant} to the current visit.

Finally, {\textsc{\textsf{{VITA}}}} obtains the patient representation $\mathbf{q}^T$ by aggregating the dense representation $\mathbf{v}^T$ of the current visit $T$ and all dense representations $\mathbf{v}^t\in\mathcal{V}^{rel}$ of the relevant past visits $t$ based on their relevance scores $\alpha^t$, as follows:
\small
\begin{equation}\label{eq:patient_representation}
    \begin{split}
    \mathbf{q}^T = \sum\nolimits_{\mathbf{v}^t\in\mathcal{V}^{rel}\cup\{\mathbf{v}^T\}}\alpha^t\mathbf{v}^t.
    \end{split}
\end{equation}
\normalsize

We highlight that {\textsc{\textsf{{VITA}}}}'s two core ideas ({\it i.e.}, relevant-visit selection and target-aware attention modules) can be orthogonally combined with the encoder of {\it any} medication recommender systems, thereby improving their accuracy; this claim will be empirically validated in Section~\ref{sec:evaluation}.

\noindent\textbf{Predictor}.
In its predictor, {\textsc{\textsf{{VITA}}}} aims to recommend the set $\hat{\mathcal{M}}^{T}$ of necessary medications for a patient based on her patient representation $\mathbf{q}^T$ obtained.
To this end, {\textsc{\textsf{{VITA}}}} first obtains, using the patient representation $\mathbf{q}^T$, the two medication representations (a) $\mathbf{p}_k^T\in\mathbb{R}^{dim}$, which captures features of the $k$-th medication that are necessary at her current visit $T$ when considering her current health status, and (b) $\mathbf{\bar{p}}_k^T\in\mathbb{R}^{|\mathcal{M}|}$, which captures the probabilities that all medications prescribed at her {\it past} visits will be recommended as the $k$-th medication at her current visit $T$ when considering her current health status. Then, {\textsc{\textsf{{VITA}}}} fuses them to obtain the $k$-th medication of the necessary medications for her current visit. The above process is iteratively performed until all necessary medications for her current visit are obtained.\footnote{In practice, {\textsc{\textsf{{VITA}}}} predicts the $<$END$>$ token (class), indicating that the medication prediction at the patient's current visit is completed. Also, we compared the accuracies of the one-by-one approach and the all-at-once approach on our framework. The results showed that the one-by-one approach was more accurate.}

\noindent\textbf{(a) Current Health-Aware Medication Representation}.
To obtain a current health-aware medication representation $\mathbf{p}_k^T$, {\textsc{\textsf{{VITA}}}} first obtains enriched medication representations $\mathbf{e}_i$ via the relations between medications, by applying an independent two-layer Graph Convolutional Network (GCN) to each of the EHR and DDI graphs with randomly initialized medication representations $\mathbf{e}_i$, and then fusing outputs of the two GCNs per medication~\cite{gamenet}.

Subsequently, {\textsc{\textsf{{VITA}}}} begins to obtain the current health-aware medication representation $\mathbf{p}_k^T$ using the patient representation $\mathbf{q}^T$ and the medication representations $\mathbf{e}_i$.
In this process, especially, {\textsc{\textsf{{VITA}}}} focuses on the medications already predicted ({\it i.e.}, medications up to the $(k-1)$\footnote{$k \geq 1$, when $k = 1$, a randomly initialized $<$START$>$ token (representation) is used.}-th) rather than all medications, since this approach helps ensure that the relations between recommended medications are considered when recommending the next one.
This process can be formally expressed as follows:
\small
\begin{equation}\label{eq:embedding representation}
    \begin{split}
    \mathbf{p}_k^T = softmax(\frac{transformer\mathbf{(e}^{T,1}_*,...,
    \mathbf{e}^{T,(k-1)}_*)\odot{\mathbf{q}^T}}{\sqrt{dim}})\mathbf{q}^T, \;\;\;\;\;\;\; \\
    \end{split}
\end{equation}
\normalsize
where $transformer\mathbf(\cdot)$ denotes a transformer-based model as used in~\cite{cognet}, which aggregates inputs considering the relation between inputs; $\mathbf{e}^{T,(k-1)}_*$ denotes a medication representation of the $(k-1)$-th predicted medication at the current visit $T$; $\odot$ denotes a dot-product.

\noindent\textbf{(b) Current Health-Relevant Past Medication Representation}.
Next, {\textsc{\textsf{{VITA}}}} obtains a current health-relevant past medication representation $\mathbf{\bar{p}}_k^T$.
To achieve this, {\textsc{\textsf{{VITA}}}} evaluates two levels ({\it spec.}, medication-level $r_{m,i}^{t}\in\mathbb{R}$ and visit-level $r_{v}^t\in\mathbb{R}$) of relevance between the current visit and each past visit~\cite{cognet}. 
Specifically, the relevance $r_{m,i}^{t}$ at the medication-level indicates how much each medication $i$ prescribed at each of the patient's past visits $t$ is related to the $k$-th medication that is necessary at her current visit $T$ ({\it i.e.}, $\mathbf{p}_k^T$), as follows:
$\forall i \in \mathbf{m}^t$; $\forall t \in \{1,\cdots, (T-1)\}$,
\small
\begin{equation}\label{eq:medication_scores}
    \begin{split}
    r_{m,i}^{t} = \frac{exp((\mathbf{e}_i^t\odot{\mathbf{p}_k^T})/\sqrt{dim})}{\sum_{j=1}^{|\mathbf{m}^t|}exp((\mathbf{e}_j^t\odot{\mathbf{p}_k^T})/\sqrt{dim})},
    \end{split}
\end{equation}
\normalsize
where $\mathbf{e}_i^t$ denotes a medication representation of medication $i$ prescribed at the $t$-th visit; $\mathbf{m}^t$ is defined in Section~\ref{sec:definition}. While the relevance $r_{v}^t$ at the visit-level indicates how much the patient's health status at each past visit $t$ is related to her current health status. So, {\textsc{\textsf{{VITA}}}} first obtains the patient representation $\mathbf{q}^t$ at the $t$-th past visit for all her past visits in the same way as in its encoder. Then, {\textsc{\textsf{{VITA}}}} employs the target-aware attention module, using the patient representation $\mathbf{q}^T$ as a query and the patient representations $\mathbf{q}^T$ and $\mathbf{q}^t$ for all her visits as keys and values (refer to Eq. (\ref{eq:relevance_score})).

Given the two levels $r_{m,i}^{t}$ and $r_{v}^t$ of relevance between the current and each past visit, {\textsc{\textsf{{VITA}}}} obtains the current health-relevant past medication representation $\mathbf{\bar{p}}_k^T$, as follows:
\small
\begin{equation}\label{eq:medication_probability_vectors}
    \begin{split}
        \mathbf{\Bar{p}}_k^T = \sum_{t=1}^{(T-1)}r_v^t\boldsymbol{r}^t_m \ \ (\boldsymbol{r}^t_m \in \mathbb{R}^{|\mathcal{M}|}), \;\;\;\;\;\;\;\;\;\;\; \\
        \text{where} \ \forall i \in \mathcal{M}, \ \boldsymbol{r}_{m}^t =
        \begin{cases}
            r_{m,i}^t \ ,& i \in \mathbf{m}^t,  \\
            \;0,  & otherwise. 
        \end{cases}
    \end{split}
\end{equation}
\normalsize

Finally, {\textsc{\textsf{{VITA}}}} fuses the two medication representations $\mathbf{p}_{k}^T$ and $\mathbf{\Bar{p}}_{k}^T$ to obtain the $k$-th medication of the necessary medications for the patient's current visit $T$, as follows:
\small
\begin{equation}\label{eq:medication_current}
    \begin{split}
    \hat{\mathcal{M}}^{T} = \hat{\mathcal{M}}^{T} \cup argmax_{k \in \{1,...,|\mathcal{M}|\}}(\hat{\mathbf{p}}_k^T),
    \;\;\;\;\;\;\;\;\;\;\;\;\;\;\;\;\;\;\\
   \text{where} \ \hat{\mathbf{p}}_k^T = \lambda_k \{ softmax(\mathbf{p}_k^T\mathbf{W}_p+\mathbf{b}_p) \} + (1- \lambda_k) \mathbf{\Bar{p}}_{k}^T, 
    \end{split}
\end{equation}
\normalsize
where the set $\hat{\mathcal{M}}^{T}$ begins as an empty set; $\hat{\mathbf{p}}_k^T \in\mathbb{R}^{|\mathcal{M}|}$ denotes the probabilities that all medications will be recommended as the $k$-th medication at her current visit $T$;
$\mathbf{W}_p\in\mathbb{R}^{dim\times|\mathcal{M}|}$ and $\mathbf{b}_p\in\mathbb{R}^{|\mathcal{M}|}$ denote a learnable weight matrix and a bias vector, respectively; $\lambda_k \in \mathbb{R}$ denotes a learnable parameter.
In other words, based on such a $\hat{\mathbf{p}}^T_k$, {\textsc{\textsf{{VITA}}}} predicts the medication with the highest probability as the $k$-th medication.

\subsection{Training}\label{sec:Training}
For the medications predicted by {\textsc{\textsf{{VITA}}}}, we employ the cross-entropy loss as the objective function, same as in~\cite{cognet}, to learn the medication representations $\mathbf{e}_i$ and other learnable parameters of {\textsc{\textsf{{VITA}}}}, as follows:
\small
\begin{equation}
    \begin{split}
        \mathcal{L}=-\sum_{t=1}^{T}\sum_{i=1}^{|\mathcal{M}|}\mathbf{m}^t_i\textit{log}(\hat{\mathbf{p}}^t_{k,i}). 
    \end{split}
\end{equation}
\normalsize

\section{Evaluation}\label{sec:evaluation}
In this section, we conducted extensive experiments aiming at answering the following key research questions (RQs):
\begin{itemize} 
\item \textbf{RQ1}: Does {\textsc{\textsf{{VITA}}}} provide more accurate recommendation than state-of-the-art medication recommenders?
\item \textbf{RQ2}: Is each of {\textsc{\textsf{{VITA}}}}'s two core ideas 
effective for medication recommendation?
\item \textbf{RQ3}: Does equipping existing methods with {\textsc{\textsf{{VITA}}}}'s two core ideas consistently improve their accuracies?
\item \textbf{RQ4}: Which past visit information does {\textsc{\textsf{{VITA}}}} select from all of the patients' past visit information?
\end{itemize}
\subsection{Experimental Settings}\label{sec:Experimental_Settings}
\textbf{Datasets}. We used the MIMIC-III dataset~\cite{mimic}, widely used in medication recommendation studies~\cite{gamenet, micron,safedrug,cognet}, and the MIMIC-IV dataset~\cite{mimiciv}, which is a follow-up dataset to MIMIC-III dataset.
Following~\cite{gamenet,micron,safedrug,cognet}, we also filtered out the patients who visited only once and the medical records related only to them. Please refer to the Appendix for detailed statistics of the two datasets.

\noindent\textbf{Competitors}. We compared {\textsc{\textsf{{VITA}}}} with six competitors: one basic classifier (Nearest as used in~\cite{gamenet}) and five state-of-the-art medication recommender systems (LEAP~\cite{leap}, GAMENet~\cite{gamenet}, MRSC~\cite{mrsc}, SafeDrug~\cite{safedrug}, and COGNet~\cite{cognet}).

\noindent\textbf{Evaluation Protocols}. We randomly split the patients in each dataset into training ($4/6$), validation ($1/6$), and test ($1/6$) sets as in~\cite{gamenet,safedrug,cognet}. To measure accuracy, as in~\cite{gamenet,sarmr,mrsc,safedrug,cognet}, we used the following three measures: Jaccard, PRAUC, and F1.
In addition, following~\cite{gamenet,sarmr,safedrug,cognet}, we used the DDI rate~\cite{gamenet} to measure the possibility of adverse effects between recommended medications. However, due to space limitations, we present only the results in terms of accuracy here. Please refer to the Appendix for the results from the DDI rate.
Furthermore, for each measure, we report the average values over five independent runs for most of our experiments; in Tables~\ref{table:sota}, \ref{table:ablation} and Figure~\ref{fig:rq3}, all improvements are statistically significant with a $p$-value $\leq 0.001$.

\subsection{Results and Analysis}\label{sec:Results_and_Analysis}
Due to space limitations, for RQ3 and RQ4, we present the results on the MIMIC-III dataset only. Please refer to the Appendix for the results from the MIMIC-IV dataset.

\noindent\textbf{RQ1: Comparison with Six Competitors}. To demonstrate the superiority of {\textsc{\textsf{{VITA}}}}, we compared the accuracy of {\textsc{\textsf{{VITA}}}} and those of the six competitors\footnote{We carefully tuned the hyperparameters of all methods. Please refer to the Appendix for the specific hyperparameter values.}.

As shown in Table~\ref{table:sota},
{\textsc{\textsf{{VITA}}}} outperforms {\it all} competitors on {\it all} datasets for {\it all} measures consistently. Specifically, on the MIMIC-III and -IV datasets, {\textsc{\textsf{{VITA}}}} outperforms the best competitor ({\it i.e.}, COGNet) by up to 3.83\% and 5.56\% in terms of Jaccard. These are {\it considerable improvements} in the sense that existing state-of-the-art medication recommender systems~\cite{gbert, sarmr, mrsc, safedrug, DrugRec, 4sdrug, cognet} tend to outperform their best competitor by an average of about 1.71\% and a maximum of about 3.00\% in terms of Jaccard.
Moreover, the improvement gap increased from 3.83\% in the MIMIC-III dataset to 5.56\% in the MIMIC-IV dataset in terms of Jaccard. This is interpreted as our idea of selecting and using only the past visits relevant to the patient's current visit becoming more important as the number of visits increases (the MIMIC-IV dataset, compared to the MIMIC-III dataset, contains about 2.15 times as many patients and about 2.53 times as many visits, with the average number of visits per patient increasing from 2.59 to 3.05).

Also, in the Appendix, we will analyze the training time of {\textsc{\textsf{{VITA}}}} and competitors for a more comprehensive understanding of the efficiency of {\textsc{\textsf{{VITA}}}}'s encoder.

\noindent\textbf{RQ2: Ablation Study}. 
In order to verify the effectiveness of our two novel ideas of {\textsc{\textsf{{VITA}}}} (relevant-visit selection and target-aware attention), we conducted comparative experiments using {\textsc{\textsf{{VITA}}}} and its variants.

\begin{table}[t!]
\centering
\small
\caption{Accuracies of six competitors and {\textsc{\textsf{{VITA}}}}. The best and the second-best results in each column ({\it i.e.}, each measure) are in bold and underlined, respectively.}\label{table:sota}
\renewcommand{\arraystretch}{1.1}
\resizebox{0.47\textwidth}{!}{
\begin{tabular}{ccccccc}
\toprule
{\textbf{Datasets}} & \multicolumn{3}{c}{\textbf{MIMIC-III}}                                   & \multicolumn{3}{c}{\textbf{MIMIC-IV}}\\ \cmidrule(lr){1-1} \cmidrule(lr){2-4} \cmidrule(lr){5-7}
\textbf{Measures}                       & \textbf{Jaccard}        & \textbf{PRAUC}     & \textbf{F1}   & \textbf{Jaccard}        & \textbf{PRAUC}     & \textbf{F1} \\ \midrule
\multicolumn{1}{c|}{\textbf{Nearest}}           & 0.3917         & 0.3813          & {0.5474}                    & {0.4523}    & {0.4461}  & {0.6045} 
\\\hline
\multicolumn{1}{c|}{\textbf{LEAP} (KDD'17)}              & 0.4306         & 0.6379          & {0.5941}                     & 0.4273        & 0.5949        & {0.5836}       \\ 
\multicolumn{1}{c|}{\textbf{GAMENet} (AAAI'19)}           & 0.4350       & 0.6754        & {0.5902}                     & 0.4600     & 0.7029        & {0.6106}      \\
\multicolumn{1}{c|}{\textbf{MRSC} (CIKM'21)}              & 0.4856       & 0.7413        & {0.6451}                   & 0.4487        & 0.6907         & {0.6063} \\
\multicolumn{1}{c|}{\textbf{SafeDrug} (IJCAI'21)}          & 0.5060         & 0.7537    & {0.6630}                    & 0.4731   & 0.6976   & {0.6251} \\
\multicolumn{1}{c|}{\textbf{COGNet} (WWW'22)}            & \underline{0.5087}         & \underline{0.7567}          & \underline{0.6666}        & \underline{0.4943}         & \underline{0.7069}          & \underline{0.6473} 
\\ \hline
\multicolumn{1}{c|}{\textbf{\textsc{\textsf{{VITA}}}}}    & \textbf{0.5282}         & \textbf{0.7673}       & \textbf{0.6815}      & \textbf{0.5218}         & \textbf{0.7148}       & \textbf{0.6685} \\ 
\bottomrule
\end{tabular}
}
\end{table}

\noindent\textbf{{\textsc{\textsf{{VITA}}}}'s variants for confirming the effectiveness of relevant-visit selection}. (1) {\textsc{\textsf{{VITA}}}}-RS does {\it not} use the relevant-visit selection module in {\textsc{\textsf{{VITA}}}}. (2) {\textsc{\textsf{{VITA}}}}-RS$_{Top\text{-}1}$ uses only one past visit, which is most similar to the current visit of a patient in terms of Jaccard similarity, and the current visit information as input for its encoder, instead of using the relevant-visit selection module. Also, we note that, when the value of temperature of the attention network is close to zero, the difference between the attention weights increases, sharpening their distribution~\cite{distill_sharp,sharp}. 
It will make our target-aware attention module have an effect similar to using only the past visits relevant to the current visit; in other words, this will be able to play a role similar to the relevant-visit selection module. Therefore, we measured the accuracies of {\textsc{\textsf{{VITA}}}}-RS, while varying the temperature $\tau_a$ of its target-aware attention module from 1 to 0.2 by 0.2 (please refer to the Appendix for detailed results of this experiment); we considered (3) {\textsc{\textsf{{VITA}}}}-RS$_{sharp}$, which is {\textsc{\textsf{{VITA}}}}-RS with the best-performing temperature $\tau_a$.

\begin{table}[t!]
\small
\centering
\caption{The effects of {\textsc{\textsf{{VITA}}}}'s two core ideas (relevant-visit selection and target-aware attention). The best result in each column ({\it i.e.}, each measure) is in bold.}\label{table:ablation}
\renewcommand{\arraystretch}{1.1}
\resizebox{0.47\textwidth}{!}{
\begin{tabular}{ccccccc}
\toprule
{\textbf{Datasets}} & \multicolumn{3}{c}{\textbf{MIMIC-III}}                                   & \multicolumn{3}{c}{\textbf{MIMIC-IV}}\\ \cmidrule(lr){1-1} \cmidrule(lr){2-4} \cmidrule(lr){5-7}
\textbf{Measures}                       & \textbf{Jaccard}        & \textbf{PRAUC}     & \textbf{F1}   & \textbf{Jaccard}        & \textbf{PRAUC}     & \textbf{F1}  \\ \midrule

\multicolumn{1}{c|}{\textbf{\textsc{\textsf{{VITA}}}}}    & \textbf{0.5282}         & \textbf{0.7673}          & \textbf{0.6815}     &\textbf{0.5218}         & \textbf{0.7148}          & \textbf{0.6685}       \\ \hline
\multicolumn{1}{c|}{\textbf{{\textsc{\textsf{{VITA}}}}-RS}}           & 0.5163         & 0.7558          & 0.6708         & 0.5197         & 0.7088          & {0.6590}        \\
\multicolumn{1}{c|}{\textbf{{\textsc{\textsf{{VITA}}}}-RS$_{Top\text{-}1}$}}       & 0.5140         & 0.7413         & 0.6702      & 0.4803        & 0.6943        & 0.6352        \\
\multicolumn{1}{c|}{\textbf{{\textsc{\textsf{{VITA}}}}-RS$_{sharp}$}}        & 0.5188       & 0.7622         & 0.6718       & 0.5199      & 0.7090         & 0.6639          \\ \hline
\multicolumn{1}{c|}{\textbf{{\textsc{\textsf{{VITA}}}}-TA$_{avg.}$}}           & 0.5119        & 0.7505         & 0.6650     & 0.5130        & 0.7069          & 0.6596         \\
\multicolumn{1}{c|}{\textbf{{\textsc{\textsf{{VITA}}}}-TA$_{RNN}$}}           & 0.5152        & 0.7538       & 0.6682    & 0.5167        & 0.7012          & 0.6633             \\
\multicolumn{1}{c|}{\textbf{{\textsc{\textsf{{VITA}}}}-TA$_{attn.}$}}           & 0.5181        & 0.7587         & 0.6699     & 0.5196       & 0.7094         & 0.6622      \\
\bottomrule
\end{tabular}
}
\end{table}

\noindent\textbf{{\textsc{\textsf{{VITA}}}}'s variants for confirming the effectiveness of target-aware attention}. (1) {\textsc{\textsf{{VITA}}}}-TA$_{avg.}$ (resp. (2) {\textsc{\textsf{{VITA}}}}-TA$_{RNN}$) employs the mean pooling (resp. a RNN-based model ({\it spec.}, GRU)) instead of the target-aware attention module in {\textsc{\textsf{{VITA}}}} when fusing the past visit information relevant to the current visit. (3) {\textsc{\textsf{{VITA}}}}-TA$_{attn.}$ employs a typical attention network that uses the current visit information as a query and the relevant past visit information as keys and values, instead of the target-aware attention module, when fusing the past visit information relevant to the current visit. 

Table~\ref{table:ablation} shows the accuracies of {\textsc{\textsf{{VITA}}}} and its all variants.

\noindent\textbf{Results regarding the effectiveness of relevant-visit selection}. We observed that {\textsc{\textsf{{VITA}}}} outperforms {\textsc{\textsf{{VITA}}}}-RS. This result indicates that using only the past visits relevant to the current visit is important in recommending effective medications to patients; in other words, the past visits irrelevant to the current visit should not be considered at all when recommending medications.
However, even though {\textsc{\textsf{{VITA}}}}-RS$_{Top\text{-}1}$ employs {\it only one} past visit most {\it relevant} to the current visit of a patient in terms of Jaccard similarity, this shows lower accuracy than {\textsc{\textsf{{VITA}}}}-RS.
This is because, although the number of past visits relevant to the current visit may differ per patient, this employs a fixed number of past visits for all patients, which has adversely affected learning by including (resp. excluding) the past visits that are actually irrelevant (resp. relevant) to the current visit. This supports the design of our relevant-visit selection module, which does not have such a restriction.
We also observed that {\textsc{\textsf{{VITA}}}}-RS$_{sharp}$ showed lower accuracy than {\textsc{\textsf{{VITA}}}}.
Note that it is difficult for {\textsc{\textsf{{VITA}}}}-RS$_{sharp}$ to flexibly select any number of past visits relevant to the current visit for each patient, because it operates similarly to the argmax function as the distribution of attention weights continues to sharpen.

\noindent\textbf{Results regarding the effectiveness of target-aware attention}. We observed that {\textsc{\textsf{{VITA}}}} outperforms all its variants related to target-aware attention ({\it i.e.}, {\textsc{\textsf{{VITA}}}}-TA$_{avg.}$, {\textsc{\textsf{{VITA}}}}-TA$_{RNN}$, and {\textsc{\textsf{{VITA}}}}-TA$_{attn.}$), confirming the effectiveness of the target-aware attention module.
Also, we observed that the accuracies for most measures improve in the order of {\textsc{\textsf{{VITA}}}}-TA$_{avg.}$, {\textsc{\textsf{{VITA}}}}-TA$_{RNN}$, and {\textsc{\textsf{{VITA}}}}-TA$_{attn.}$. The results indicate that (1) capturing the degree of relevance between the current visit and each past visit plays an important role in obtaining higher accuracy (this is from the fact the accuracy of {\textsc{\textsf{{VITA}}}}-TA$_{avg.}$ is worse than those of {\textsc{\textsf{{VITA}}}}-TA$_{RNN}$, {\textsc{\textsf{{VITA}}}}-TA$_{attn.}$, and {\textsc{\textsf{{VITA}}}} in most measures); (2) such relevance may not be consistent with the order of the visits (this is from the fact the accuracy of {\textsc{\textsf{{VITA}}}}-TA$_{RNN}$ is worse than those of {\textsc{\textsf{{VITA}}}}-TA$_{attn.}$ and {\textsc{\textsf{{VITA}}}} in most measures); (3) it is important to accurately capture the degree of relevance between the current visit and each past visit in providing better accuracy (this is from the fact the accuracy of {\textsc{\textsf{{VITA}}}}-TA$_{attn.}$ is worse than that of {\textsc{\textsf{{VITA}}}} in all measures). These findings again support the limitation of existing works that we previously pointed out.

\begin{figure}[t]
\centering
\includegraphics[width=0.47\textwidth]{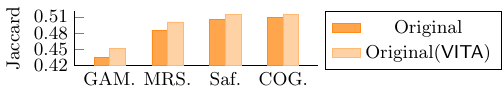}
\caption{Accuracies of four longitudinal-based methods and their variants equipped with two core ideas of {\textsc{\textsf{{VITA}}}}.}\label{fig:rq3}
\end{figure}

Also, it bears mentioning that all variants of {\textsc{\textsf{{VITA}}}}, except for {\textsc{\textsf{{VITA}}}}-RS$_{Top\text{-}1}$, outperform the {\it best competitor} ({\it i.e.}, COGNet) for most measures (compare Tables~\ref{table:sota} and~\ref{table:ablation}), even though they incorporate {\it only} one of {\textsc{\textsf{{VITA}}}}’s two core ideas; this validates that {\textsc{\textsf{{VITA}}}}'s single core idea alone is beneficial in improving the accuracy of medication recommendation.

\noindent\textbf{RQ3: Compatibility of {\textsc{\textsf{{VITA}}}}'s Two Core Ideas}.
In Section~\ref{sec:key_component}, we claimed that {\textsc{\textsf{{VITA}}}}'s two core ideas can be applied orthogonally to most medication recommenders, thereby improving their accuracy. To validate this claim, we compare the accuracy of four longitudinal-based methods (GAMENet, MRSC, SafeDrug, and COGNet) and their variants, each of which is equipped with the two core ideas of {\textsc{\textsf{{VITA}}}} (GAMENet({\textsc{\textsf{{VITA}}}}), MRSC({\textsc{\textsf{{VITA}}}}), SafeDrug({\textsc{\textsf{{VITA}}}}), and COGNet({\textsc{\textsf{{VITA}}}})); the results of the methods are represented as the first three letters of the methods in Figure~\ref{fig:rq3}.

As shown in Figure~\ref{fig:rq3}, the variants of existing methods equipped with the two core ideas of {\textsc{\textsf{{VITA}}}} outperform the original methods in Jaccard (please refer to the Appendix for the results from PRAUC and F1); that is, the result shows that our two core ideas are equipped orthogonally to most medication recommenders, improving their accuracy.

\begin{figure}[t]
\centering
\includegraphics[width=0.47\textwidth]{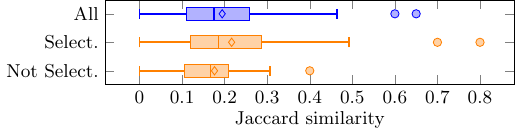}
\caption{The statistics of the Jaccard similarity between patients' current visit and `all' past visits, (i) (resp. (ii)) the past visits `selected' (resp. `not selected') by the relevant-visit selection module.}\label{fig:rq4}
\end{figure}

\noindent\textbf{RQ4: Analysis of Selected Past Visits}. One of the two core ideas of {\textsc{\textsf{{VITA}}}}, the relevant-visit selection, automatically selects only the past visit relevant to the current visit. To investigate which past visit of patients was selected by the relevant-visit selection module, we analyze the visit information of patients in the test set. We first calculated the Jaccard similarity between the current and each past visit information of a patient in the test set as in Section~\ref{sec: motivation}.
Then, we divided the past visits of patients into the following two groups: (i) (resp. (ii)) the past visits selected (resp. not selected) by the relevant-visit selection module; the statistics of the Jaccard similarity for each group, along with those of all past visits for comparison, are represented in Figure~\ref{fig:rq4}.

As shown in Figure~\ref{fig:rq4}, the relevant-visit selection module tended to select the past visit information with a high Jaccard similarity to the current visit information. Numerically, the average of Jaccard similarity between the past visit information selected by the relevant-visit selection module and the current visit information is 0.2164, which is 22.81\% higher than that for the past visit not selected by the relevant-visit selection module (0.1762). Note that the relevant-visit selection module did not always select the past visit information similar to the current one of the patients in Jaccard similarity.
This implies that, even if the past visit information is not similar to the current one in Jaccard similarity ({\it i.e.}, on the face of it), it can be useful for accurate medication recommendation ({\it e.g.}, due to its possible inherent relevance with the current one), as inferred from the superior performance of {\textsc{\textsf{{VITA}}}}, which is equipped with the relevant-visit selection.
Additionally, a case study on past visits dissimilar to the current visit in Jaccard similarity but selected by the relevant-visit selection module would be shown in the Appendix.

\section{Conclusions}\label{sec:conclusions}
In this paper, we demonstrated the following two important points w.r.t. representing a patient’s current health
status: (1) it is required to accurately capture the relevance between current visit information and each of past visit information; (2) using past visit information irrelevant to the current one is harmful;
in other words, only the past visits relevant to a patient’s current visit should be `carefully chosen', and the chosen past visits should be `weighted less' when aggregated according to the degree of relevance.
Considering these points, we proposed a novel medication recommendation framework, named {\textsc{\textsf{{VITA}}}}, based on the following two core ideas: {\it relevant-visit selection}, which allows for flexible selection of only the past visits relevant to the current visit per patient (even allowing for the selection of {\it all} past visits or selecting {\it none} of them, if necessary);
{\it target-aware attention}, which accurately captures the relevance score between a patient's current visit and each of past visits (even covering when {\it all} the past visits are {\it weakly} relevant to the current visit).
Our two novel ideas are effective in improving the accuracy, thereby making {\textsc{\textsf{{VITA}}}}, which is the final version equipped with all the ideas, consistently and significantly more accurate than its six competitors on real-world datasets. Furthermore, our two core ideas can be applied to the various medication recommender systems (even in the recommender systems of another domain) orthogonally.

\section{Acknowledgments}
This work was supported by Culture, Sports and Tourism R\&D Program through the Korea Creative Content Agency grant funded by the Ministry of Culture, Sports and Tourism in 2023 (Project Name: Development of Intelligent Personalized Rehabilitation Service Technology, Project Number: SR202104001, Contribution Rate: 33.34\%) and by Institute of Information \& Communications Technology Planning \& Evaluation (IITP) grant funded by the Korea government (MSIT) (No. 2022-0-00352 and No. RS-2022-00155586).

\bibliography{aaai24}
\end{document}